\documentclass[a4paper,11pt]{article}
\usepackage{jcappub} 

\usepackage{lineno}
\usepackage{orcidlink}

\arxivnumber{2604.06704} 
\title{Biases in the Determination of Correlations Between Underground Muon Flux and Atmospheric Temperature}

\author[a]{Bangzheng~Ma\orcidlink{0000-0002-4340-4238},}
\author[b]{Katherine~Dugas\orcidlink{0000-0001-5112-2241},}
\author[c,d,e]{Kam-Biu Luk\orcidlink{0000-0002-1694-3172},}
\author[b]{Juan~Pedro~Ochoa-Ricoux\orcidlink{0000-0001-7376-5555},}
\author[f]{Bed\v{r}ich~Roskovec\orcidlink{0000-0003-0660-5951}}
\author[a]{and~Qun~Wu\orcidlink{0000-0003-1498-068X}}
\affiliation[a]{School of Physics, Shandong University, Jinan, Shandong, China}
\affiliation[b]{Department of Physics and Astronomy, University of California, Irvine, California, USA}
\affiliation[c]{Jockey Club Institute for Advanced Study and Department of Physics, The Hong Kong University of Science and Technology, Clear Water Bay, Hong Kong, China}
\affiliation[d]{Department of Physics, University of California, Berkeley, California, USA}
\affiliation[e]{Lawrence Berkeley National Laboratory, Berkeley, California, USA}
\affiliation[f]{Faculty of Mathematics and Physics, Charles University, Prague, Czech Republic}

\emailAdd{mabangzheng@mail.edu.cn}
\emailAdd{kvdugas@uci.edu}
\emailAdd{k\_luk@berkeley.edu}
\emailAdd{jpochoa@uci.edu}
\emailAdd{bedrich.roskovec@matfyz.cuni.cz}
\emailAdd{wuq@sdu.edu.cn}

\abstract{The underground rates of cosmic-ray muons exhibit seasonal variations correlated with effective atmospheric temperature, quantified via a single coefficient. We compare two analysis methods
for studying the correlation: the standard Unbinned Method, where all rate-temperature data points are fit simultaneously via linear regression, and the Binned Method, where data points with similar temperatures are first grouped into bins before fitting. We find that while both methods are unbiased 
in the limit of negligible temperature uncertainties,
the Binned Method develops significant bias when
temperature uncertainties are present, due to binning-induced distortions. In
contrast, the Unbinned Method remains robust if the uncertainties are accurately
known. To address the widely encountered issue of imprecise uncertainty  
estimation, we
propose a novel procedure that assesses correlation stability by varying 
the time intervals and their assigned uncertainties. This approach resolves
methodological
tensions in  
studies of seasonal modulation 
of the muon rate and provides a practical framework
for robust correlation estimation under real-world conditions.}
  
\begin{document}
\maketitle
\flushbottom

\section{Introduction}

Cosmic-ray muons originate from 
 the decays of mesons that are produced 
in the interactions between  energetic cosmic rays and air molecules.
 Underground detectors observe a measurable seasonal modulation in the muon rate, driven primarily by fluctuations in
 the atmospheric temperature.
As  the atmospheric temperature increases, the reduced air density lowers the probability 
of mesons  
interacting with air molecules. 
Consequently, more mesons survive and subsequently decay to muons, leading to a positive correlation between the underground muon 
rate and atmospheric temperature. The magnitude of this correlation depends on the overburden,  
as the overlying rock preferentially attenuates low-energy muons.

\begin{figure}[!ht]
  \centering
        \centering
         \includegraphics[width=.46\textwidth]{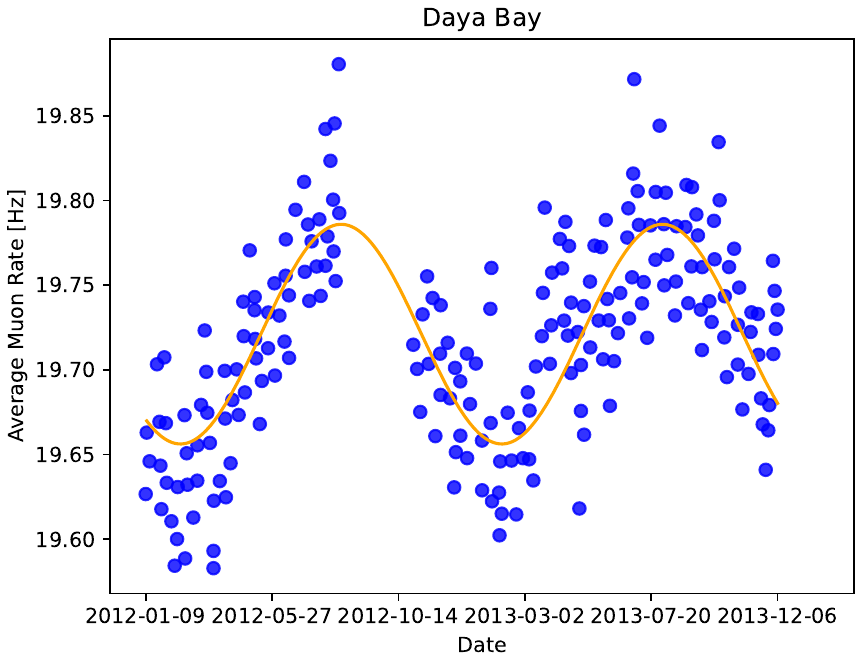}
        \centering
         \includegraphics[width=.46\textwidth]{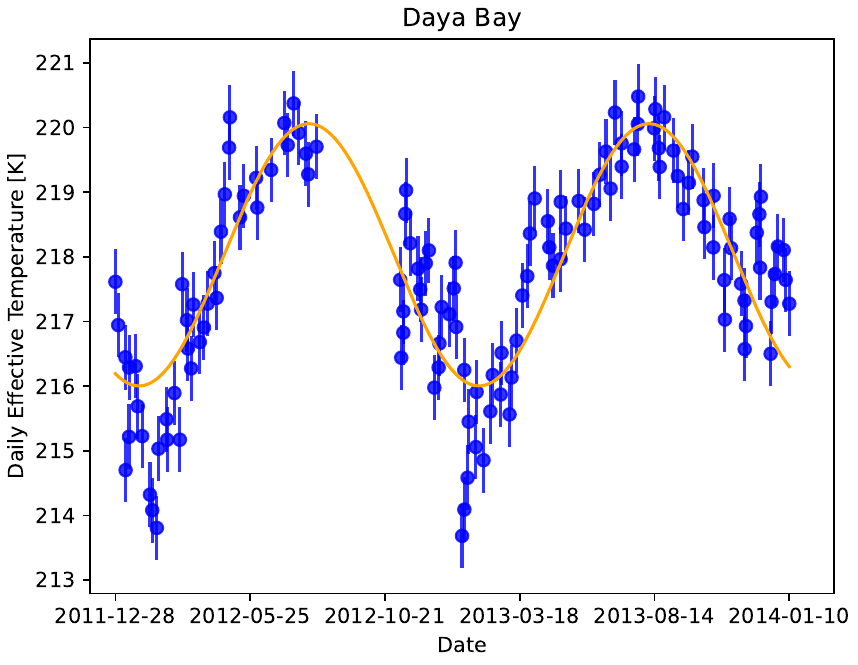}
        \centering
         \includegraphics[width=.46\textwidth]{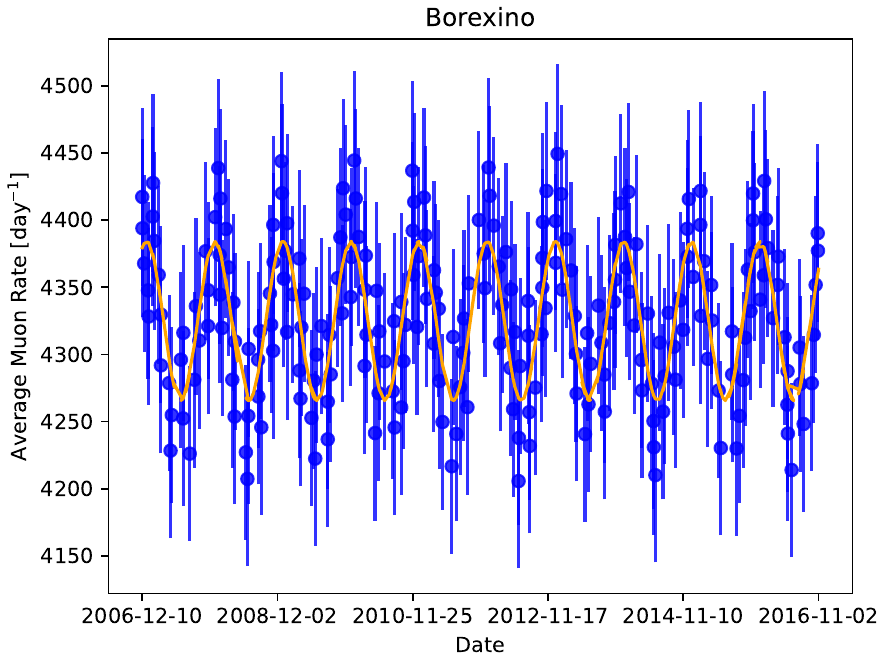}
        \centering
         \includegraphics[width=.46\textwidth]{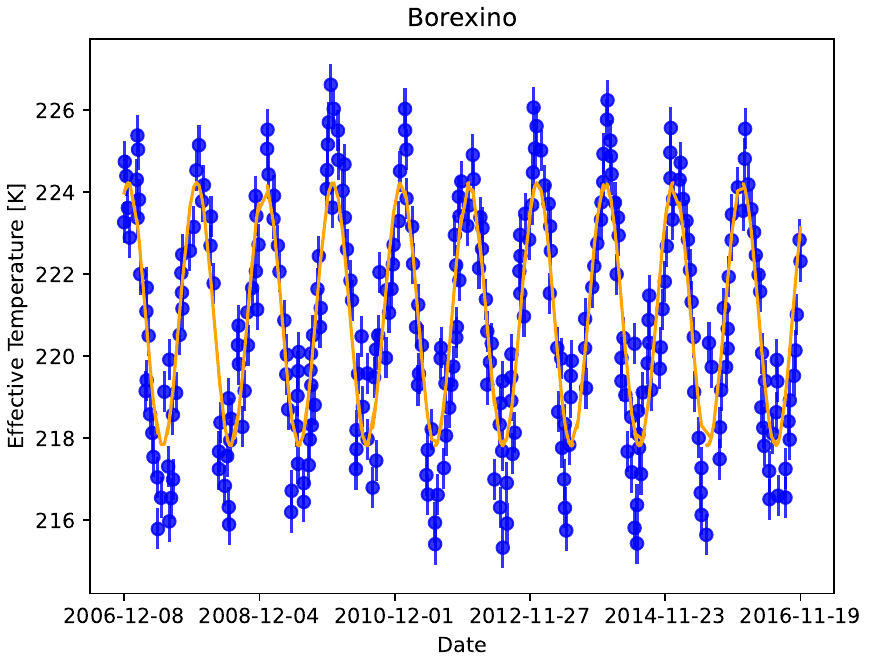}
    \caption{
    Temporal evolution of the muon rate and effective atmospheric temperature as observed by antineutrino detector module AD2 in the near experimental hall (EH1) of the Daya Bay experiment~\cite{DayaBay} (top panels) and by the  Borexino experiment~\cite{Borexino:2018pev} (bottom panels). The overlaid curves
    are drawn   using a sinusoidal function with parameters 
    obtained from fits to the experimental data reported in Refs.\cite{DayaBay,Borexino:2018pev}. Panels are reproduced from Refs.~\cite{DayaBay,Borexino:2018pev} with minor adaptations.}
    \label{fig:DYB_example}
\end{figure}

The seasonal variation of the underground muon rate is commonly 
described as linearly correlated with variations in  the effective atmospheric temperature ($T_\mathrm{eff}$)~\cite{Barrett:1952woo}. The latter is obtained by modeling the atmosphere as 
a thermal system consisting of
isothermal layers, each with a different temperature and a weight proportional to its contribution to muon production.
Theoretically, the determination of these weights comes from solutions to   cascade equations~\cite{GAISSER} that describe atmospheric shower development.
This approach has evolved from early analytical approximations considering only the pion contribution~\cite{Barrett:1952woo} to more unified forms incorporating both pion and kaon terms~\cite{Theoretical}. More recent developments employ numerical solutions, yielding formulations tailored to specific detector configurations~\cite{Theory_2024}. 
The correlation is formulated as: 
\begin{equation}
\label{eqn}
\Delta R = \alpha\,\Delta T_{\mathrm{eff}},
\end{equation}
where $\Delta R\equiv\frac{R-\langle R\rangle}{\langle R\rangle}$ is the relative difference in the muon rate $R$ at a certain time with respect to the average muon rate $\left<R\right>$ and $\Delta T_{\mathrm{eff}} \equiv \frac{T_{\mathrm{eff}}-\langle T_{\mathrm{eff}}\rangle}{\langle T_{\mathrm{eff}}\rangle}$ represents the relative difference in the effective temperature $T_\mathrm{eff}$ at a 
given time with respect to the average effective temperature $\left<T_\mathrm{eff}\right>$ over a 
period of time. 
The parameter $\alpha$ is the proportionality constant or the correlation coefficient.
This linear correlation has been observed by numerous experiments~\citep{MACRO:1997teb,MINOSfar,Borexino:2012wej,MINOSnear,DoubleC,DayaBay,Borexino:2018pev,LVD,OPERA:2018jif,COSINE-100:2021} at different 
overburdens. Figure~\ref{fig:DYB_example} shows an example of the temporal evolution of the muon rate and the effective temperature 
from two experiments, Daya Bay~\cite{DayaBay} and Borexino~\cite{Borexino:2018pev}.

Once the relative muon rate and effective temperature are determined for a set of time bins, 
linear regression is performed on the resulting data points to extract the correlation coefficient between the two quantities. We refer to 
this approach as the \textbf{“Unbinned Method”}. To minimize 
the statistical uncertainties, muon rates must be determined from sufficiently large samples. Accordingly, experiments adopt different 
binning in time based on their observed rates. For example, the MACRO experiment used monthly time bins (one data point per month)~\cite{MACRO:1997teb}, the Daya Bay experiment employed daily bins~\cite{DayaBay}, and MINOS binned the near-detector  
data in 6-hour intervals ~\cite{MINOSnear}. The corresponding effective temperatures were calculated over the same time intervals.  

 An alternative technique was introduced by the MINOS collaboration~\cite{MINOSnear}. In this 
 \textbf{"Binned Method"}, data points  
 with similar effective temperatures are grouped into discrete $T_\mathrm{eff}$ bins; for example, MINOS used bins with  
 width of 1~K. The muon rate  
 of each bin is calculated by dividing the total number of events by the cumulative livetime of all 
 the data  
 included in that bin.  
 MINOS implemented 
 both methods in their analysis. 
 The Binned Method produced a correlation coefficient  
 lower than that of the Unbinned Method. The difference between the two results was taken as a systematic uncertainty.

 With the advent of increasingly precise measurements of the underground muon rate from modern experiments and higher-quality atmospheric temperature datasets, there is growing potential to probe subtle features of the correlation between 
 the underground muon  
 rate and 
 the atmospheric temperature, including 
 possible deviations from the expected behavior.   
 However, it is essential to ensure that the analysis methodology itself does not introduce spurious biases or artifacts. To this end, we have conducted
a systematic investigation using simulation to rigorously compare the two commonly used approaches, identify their inherent limitations, and clarify the origin of the observed discrepancy reported by MINOS.
 In Section~\ref{sec-UnknownTeff}, we establish the methodological foundation of our regression analysis. 
 Our 
 simulation framework is described in Section~\ref{sec-toyMC-setup}. Section~\ref{sec-linear} presents a comparative evaluation of the two methods under varying levels of uncertainty in the effective temperature. Section~\ref{sec:influence} examines scenarios in which the uncertainties in the effective temperature 
 are inadequately estimated. In Section~\ref{sec:better_estimation}, we propose a method to mitigate 
the bias introduced by the improper error assignments to the effective temperature. Finally, Section~\ref{sec:conclusion} provides a detailed discussion of the implications of our findings.

\section{Regression procedure}\label{sec-UnknownTeff}

Given that measurement errors are inherent in both the muon rate and the effective temperature, we employ 
the Weighted Total Least Squares (WTLS) method~\cite{WTLS} 
with the $\chi^2$ expression defined by

\begin{equation}
\label{eq:chisq}
\chi^{2}=\sum_{i}\frac{\left(y_{i}-\alpha x_{i}\right)^{2}}{\sigma^{2}_{y_{i}}+\alpha^{2}\sigma^{2}_{x_{i}}},
\end{equation}
where the index $i$ runs over all data points, $x_{i}$ and $y_{i}$ correspond to the 
$i$-th $\Delta T_{\mathrm{eff}}$ and $\Delta R$ 
values derived from Equation~\ref{eqn}, respectively, and $\sigma_{x_{i}}$ and $\sigma_{y_{i}}$ represent the corresponding errors. The best-fit value of $\alpha$ is obtained by
minimizing the $\chi^2$ using TMinuit~\cite{ref-minuit,web-minuit}. 
To validate our implementation, we cross checked the results using the Weighted Orthogonal Distance Regression 
algorithm implemented in scipy.odr~\cite{scipy_odr}. This algorithm numerically minimizes the sum of the weighted orthogonal distances between
the data points and the fitted line.
Note that the uncertainties entering the regression are those 
of the individual measurements, whereas  
systematic uncertainties related to 
the muon rate and effective temperature are typically correlated between
measurements and do not enter the regression directly. 
 
Determining the statistical uncertainty of the individual muon rate  
is straightforward: muon counts  
in a given time interval follow Poisson statistics, yielding a well-defined standard deviation. 
In contrast, quantifying the uncertainty of the effective temperature  
is considerably more challenging. The effective temperature is calculated from 
the profiles of the atmospheric temperature,  
weighted to  account for the altitude dependence of muon production.
Historical approaches 
for estimating its uncertainty vary significantly.  The MACRO experiment~\cite{MACRO:1997teb} used  
balloon-borne measurements from the Italian Aeronautics Authority to obtain  
the local temperature profiles 
directly at specific atmospheric depths between 1991 and 1994. For each monthly time bin, the effective temperature was calculated as the mean of the instantaneous effective temperatures within that interval, and the standard deviation was taken as a statistical uncertainty. The MINOS collaboration~\cite{MINOSfar} cross-validated the effective temperatures derived from the ERA-Interim~\cite{ERA-Interim} reanalysis against an independent data set from the Integrated Global Radiosonde Archive (IGRA)~\cite{IGRA}. The daily differences between the ERA-Interim- and IGRA-based effective temperatures were well modeled by a Gaussian distribution, and their spread was used as a bin-to-bin uncorrelated uncertainty. The Daya Bay experiment~\cite{DayaBay} followed the same ERA-Interim–IGRA comparison approach and, in addition, 
propagated 
the 
uncertainties associated with the temperature weights used in computing the effective temperature.
Another issue is that the weights used in calculating the effective temperature  
depend on which  
theoretical model 
is employed~\cite{Theory_2024}.

Given the fact that there are several approaches used to estimate the uncertainty of the effective temperature, it is important 
to assess the reliability and achievable accuracy of the WTLS in 
the study of seasonal modulation of cosmic-ray muons. 
To this end, we performed a systematic investigation of 
the errors-in-variables regression under controlled conditions, where the uncertainties in the  
independent variable $\Delta T_\mathrm{eff}$ are imperfectly known, using dedicated toy Monte Carlo simulations.

\section{Monte Carlo simulation}\label{sec-toyMC-setup}

\begin{figure}[htbp]
\centering
\includegraphics[width=.45\textwidth]{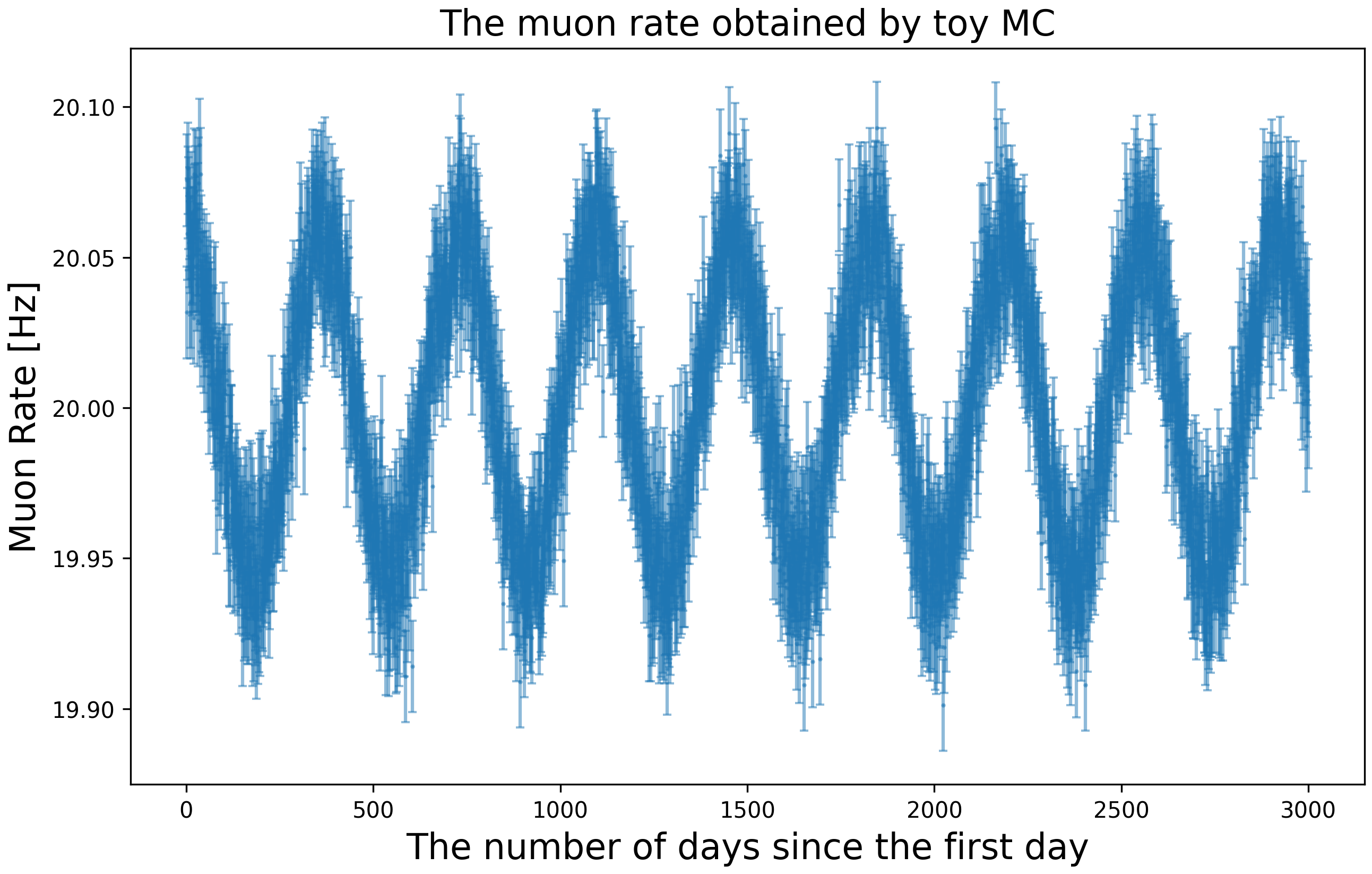}\label{fig:MC_muonrate}
\qquad
\includegraphics[width=.45\textwidth]{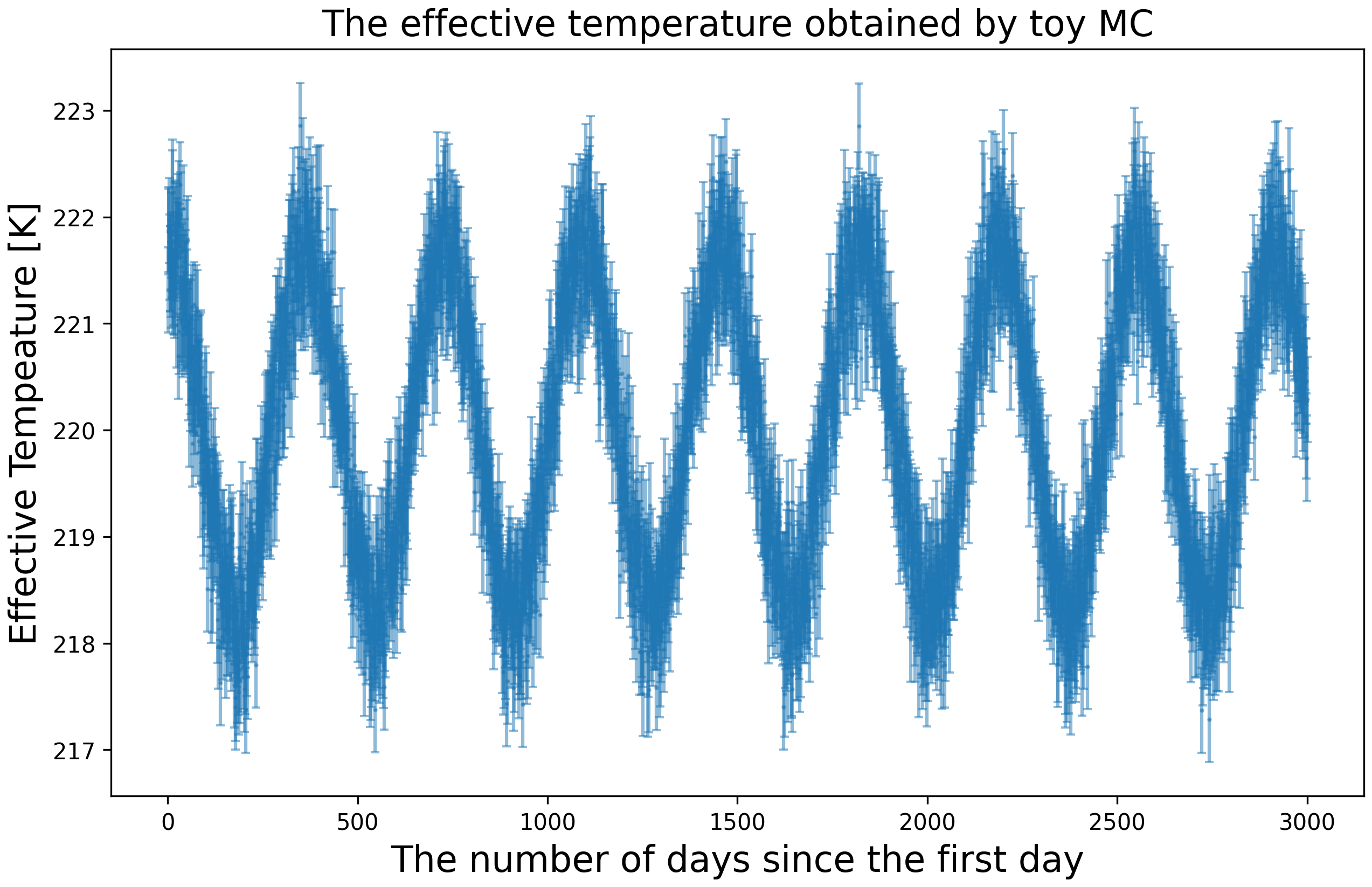}\label{fig:MC_temperature}
\caption{
    Example of the temporal
    evolution of the simulated muon rate and effective temperature over a period of 3000 days.  
    The random uncertainty
    in the muon rate at 
    the $i$-th day is given by $\sqrt{\left(R_{i}/86400\right)}$, where $R_i$ is the muon rate in Hz. The random 
    uncertainty 
    in
    the effective temperature is fixed at 0.4~K for all data points. \label{fig:MC_history}}
\end{figure}

To investigate potential biases under controlled conditions while mimicking experimental data, we created a 
toy Monte Carlo (toyMC) data set with daily time bins.
To do so, we assume a strict linear relationship between $\Delta R$ and $\Delta T_\mathrm{eff}$. 
Both the daily muon rate $R$ and the daily effective temperature $T_\mathrm{eff}$ 
follow periodic functions
with identical period. Independent random measurement errors 
are also incorporated in
each data  
point.
Formally, we define the measured daily muon rate (or effective temperature) $Y(t_i)$ as:
    \begin{equation}
        Y(t_i)=Y_{0}\left(1+A\cdot \cos\left[\frac{2\pi}{P}(t_i-t_{0})\right]\right)+\delta_i \label{eq:MCset}
    \end{equation}
where:

\begin{itemize}
\item $Y_{0}$ is the time-averaged value of the muon rate $\left< R \right>$ or the effective temperature $\left< T_\mathrm{eff} \right>$  over the entire observation period;
\item $A$ is the amplitude of the periodic variation (dimensionless relative amplitude);
\item $P$ is the  
period of the oscillation;
\item $t_0$ is the  
time
offset (time of peak amplitude) in units of days; 
\item $t_{i}$ is  
given by 
$t_i = i + t_0$ for the $i$-th day;
\item $\delta_i \sim \mathcal{N}(0,\sigma^2)$ 
is the measurement error
drawn from a normal distribution with mean 0 and standard deviation $\sigma$. 
\end{itemize}
The functional form of 
 Equation~\ref{eq:MCset} is inspired by experimental observations in Daya Bay~\cite{DayaBay}, and  
the parameter values  based on  those from Daya Bay are detailed in Table~\ref{table:para_MC}. The values of the amplitudes $A_{T_\mathrm{eff}}$ and $A_{R}$ yield a true value of the correlation coefficient $\alpha_\mathrm{true}=\frac{A_R}{A_{T_\mathrm{eff}}}\doteq0.359$. 
 Figure~\ref{fig:MC_history} shows an example of the simulated daily muon rate for a 3000-day observation period, with statistical uncertainty $\sigma_i = \sqrt{R_i/86400}$. Here, $R_i$ is expressed in Hz, assuming 24-hour continuous data collection each day. The figure also shows the effective atmospheric temperature toy data set, with daily values fluctuating according to a bin-to-bin uncorrelated uncertainty $\sigma_\mathrm{T} = 0.4$~K.

    \begin{table}[ht]
        \centering
        \begin{tabular}{|c|c|c|c|c|}
        \hline
          & $Y_{0}$ & $A$ & $P$ (Days)& $t_{0}$ \\
        \hline
        Muon Rate & 20 Hz & 0.0028 & 365 & 0 \\
        \hline
        Effective Temperature & 220 K & 0.0078 & 365 & 0 \\
        \hline
        \end{tabular}
        \caption{Values of the parameters in Equation~\ref{eq:MCset} used
        for the muon rate and the effective temperature simulation.}
        \label{table:para_MC}
    \end{table}

\section{Unbinned and Binned Methods}
\label{sec-linear}
 
The daily relative variation in muon rate $\Delta R$ and effective temperature $\Delta T_\mathrm{eff}$ were calculated for the data set described in 
Section~\ref{sec-toyMC-setup}.
For the Unbinned Method, the daily data points were used directly in the linear regression utilizing the WTLS method. For the Binned Method, these points were binned into intervals of 0.2\%, covering the range [-1.5\%, 1.5\%] in $\Delta T_\mathrm{eff}$. Within each interval, we calculated the mean muon rate and the mean $\Delta T_\mathrm{eff}$, and then these averages were used as data points for linear regression.

\begin{figure}[htbp]
\centering
\includegraphics[width=.45\textwidth]{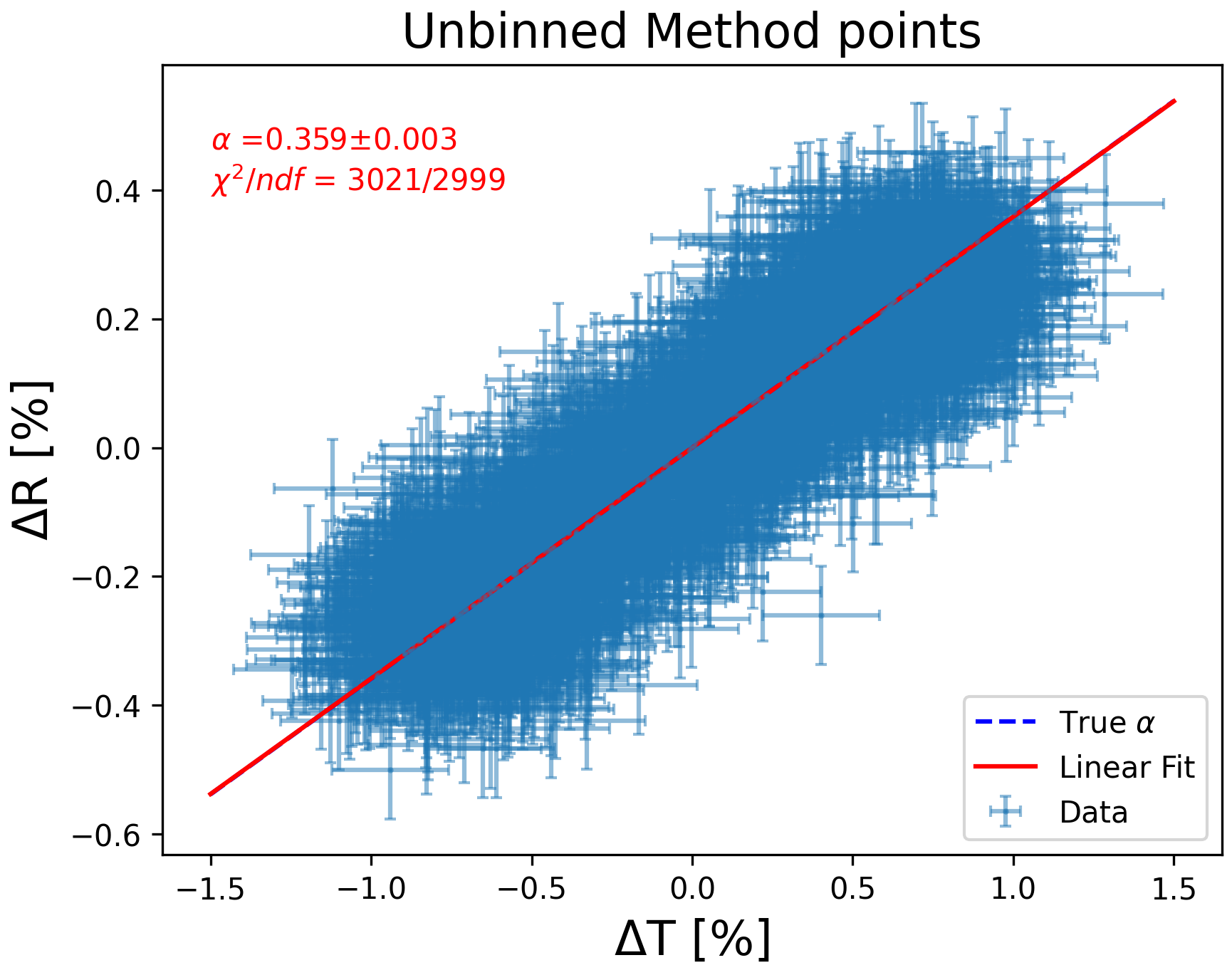}\label{fig:MC_Fit_linear_scatter}
\qquad
\includegraphics[width=.45\textwidth]{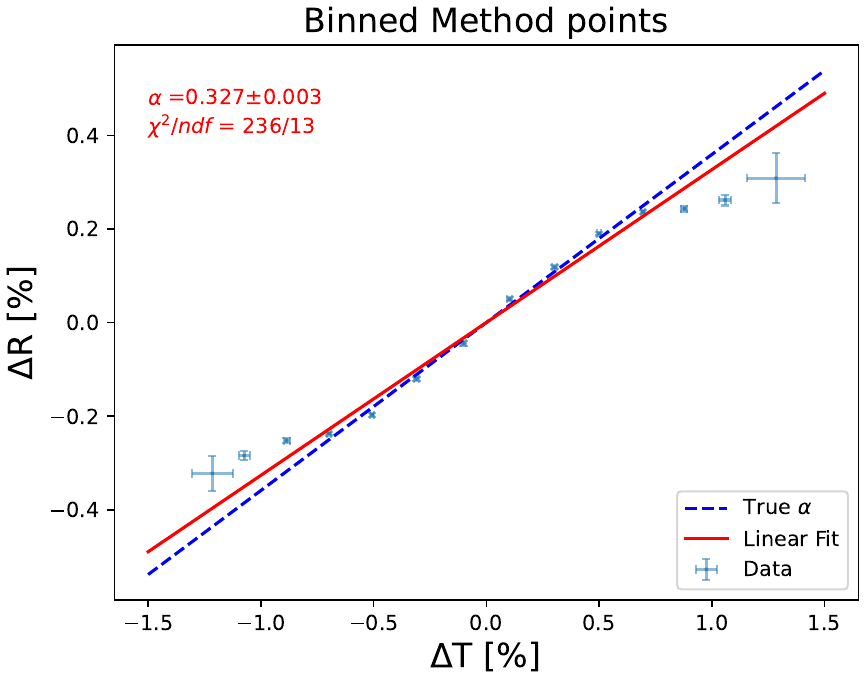}\label{fig:MC_Fit_linear_binned}
\caption{Linear fits to a single toyMC data set using the Unbinned Method (left) and the Binned Method (right).\label{fig:MC_Fit_linear}}
\end{figure}

Figure~\ref{fig:MC_Fit_linear} shows the results of the regression for both methods. The Unbinned Method yields a best-fit value of $0.359 \pm 0.003$ for the correlation coefficient.
This result is consistent with the true value of $0.359$ within uncertainties, demonstrating that the estimator is unbiased.
In contrast, the binned data exhibit an intrinsic “S-shaped” distortion, causing the linear fit to systematically underestimate the
correlation coefficient.
In this case, the best-fit value is
$\alpha = 0.327 \pm 0.003$.

\begin{figure}[htbp]
\centering
\includegraphics[width=.45\textwidth]{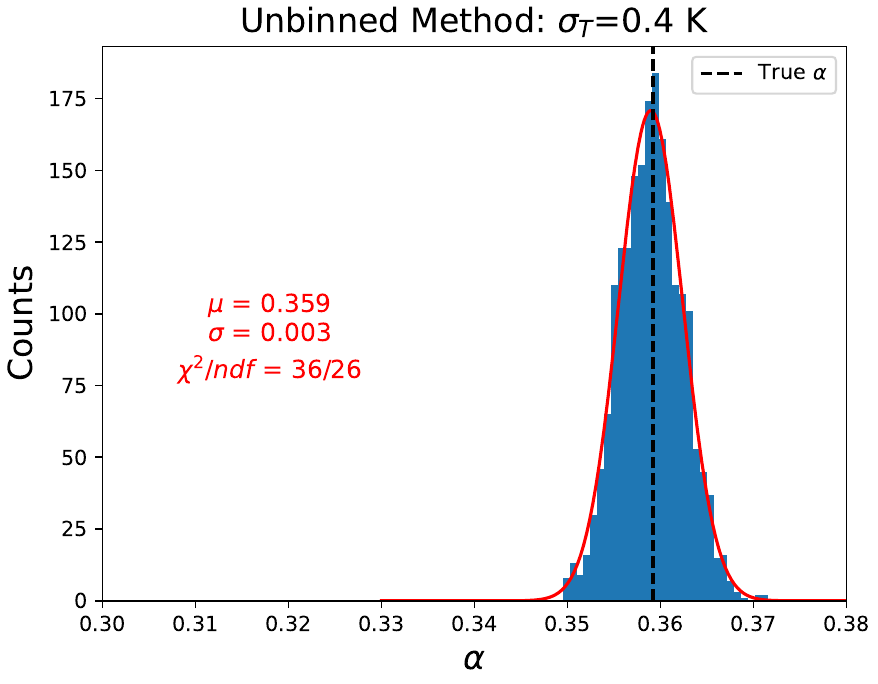}\label{fig:MC_result_scatter_linear}
\qquad
\includegraphics[width=.45\textwidth]{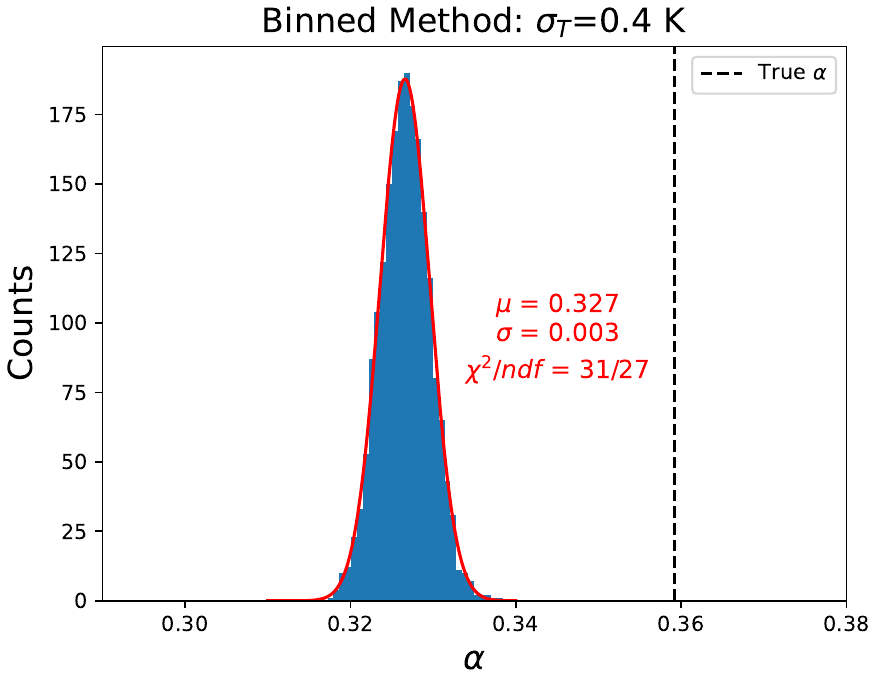}\label{fig:MC_result_binned_linear}
\caption{Distribution of correlation coefficients derived from 2,000 toyMC data sets for the Unbinned Method (left) and the Binned Method (right). The mean of the Gaussian fit is taken as the central value of $\alpha$, with $\sigma$ 
representing the uncertainty
.\label{fig:MC_result_linear}}
\end{figure}

To quantify the reliability of the analysis, we repeated this procedure with 2,000 independent toyMC data sets. In each sample, we applied both the Unbinned Method and the Binned
Method to estimate the correlation coefficient $\alpha$.  The resulting  distributions of $\alpha$ for both methods are
shown in Figure~\ref{fig:MC_result_linear}, where Gaussian fits are applied to obtain  the central values (mean)  and their uncertainties. The Unbinned Method yields $\alpha_\mathrm{unbinned} = 0.359 \pm 0.003$. This result is consistent with the input value of  $\alpha_\mathrm{true}=0.359$. In contrast, the Binned Method yields $\alpha_\mathrm{binned} = 0.327 \pm 0.003$, a
significant deviation from $\alpha_\mathrm{true}$. This discrepancy arises from the spurious S-shaped trend introduced
by binning, which biases the fit and distorts the slope parameter.

\begin{figure}[htbp]
\centering
\includegraphics[width=.45\textwidth]{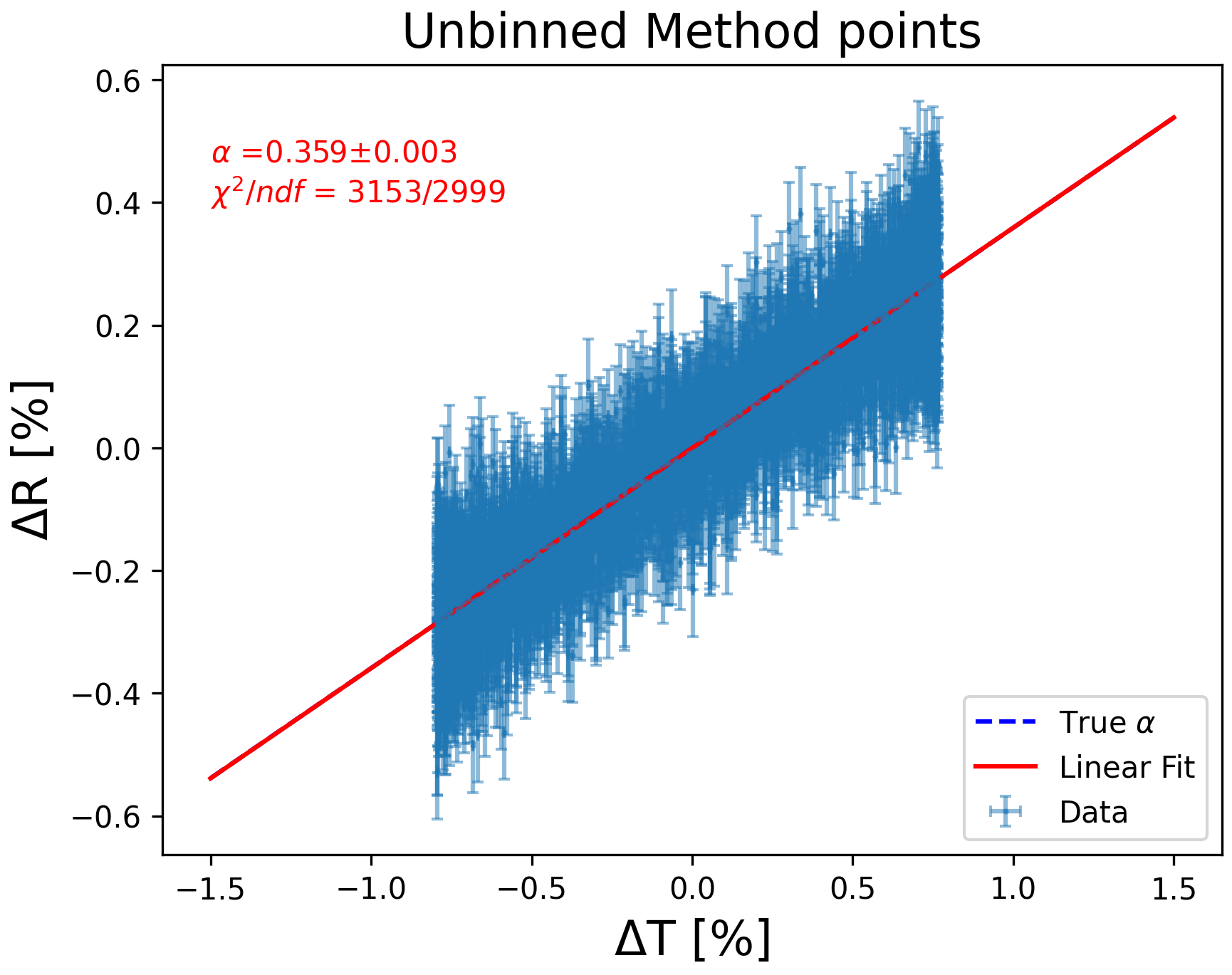}\label{fig:MC_Fit_scatter_0ET}
\qquad
\includegraphics[width=.45\textwidth]{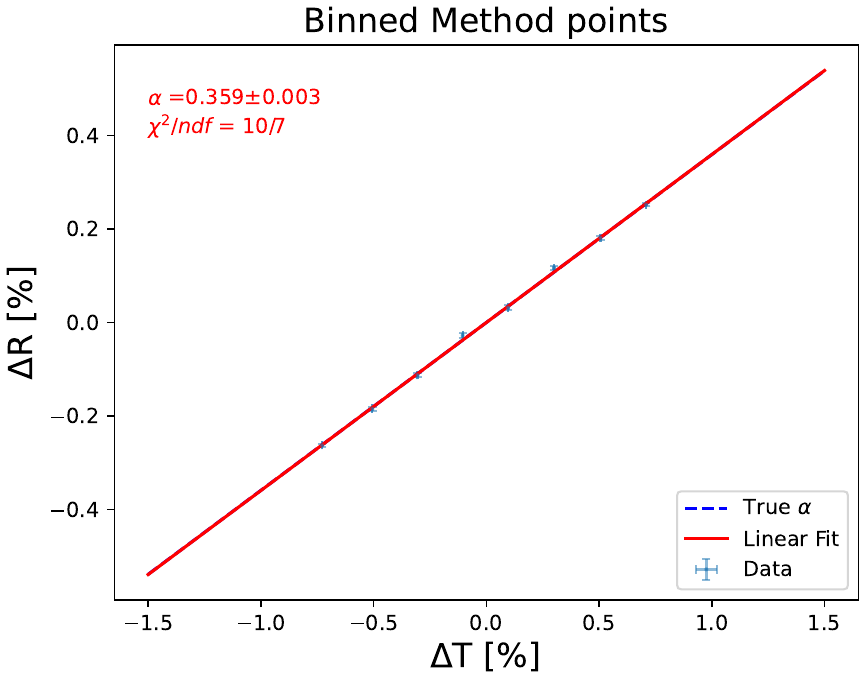}\label{fig:MC_Fit_binned_0ET}
\caption{Linear  
fits  
with
the Unbinned and Binned Methods using a toyMC data 
set 
with perfect 
knowledge of temperature 
($\sigma_T^{true}$=0).
\label{fig:MC_Fit_0ET}}
\end{figure}

To elucidate the origin of the spurious S-shaped trend in the binned data, we conducted toyMC simulations 
without any error in temperature, that is, 
$\sigma_\mathrm{T} = 0$.  
 As shown in Figure~\ref{fig:MC_Fit_0ET}, the S-shaped distortion vanishes entirely in this case: 
 each point  
 has only 
 an
 uncertainty in the 
 relative muon rate (y-direction), 
 with all points tightly clustered around the true linear relationship.  
 A non-zero 
uncertainty 
in the effective temperature translates 
the data 
point
along the $ \Delta T_{\mathrm{eff}} $ axis (x-direction), distorting the underlying correlation. 
 This translation
 biases the binned averages of the relationship $\Delta R {\mathrm{-}} \Delta T_\mathrm{eff}$ by mixing data points with different true values within each 
 bin:
 \begin{itemize}
 \item{For $\Delta T_\mathrm{eff} < 0$:  
 the translation preferentially mixes  
 points with higher true 
 $\Delta R$, leading to a net upward shift in the binned average. }
 \item{For $\Delta T_\mathrm{eff} > 0$:  
 the translation preferentially mixes 
 points with lower true $\Delta R$, leading to a net downward shift in the binned average.}
 \end{itemize}
Although statistical fluctuations in  
small data samples can cause individual bins, particularly those near $\Delta T_\mathrm{eff} \approx 0$, to deviate from this trend, the systematic bias increases with $|\Delta T_\mathrm{eff}|$, producing the characteristic S-shaped distortion, consistent with the expected behavior of regression dilution in grouped data~\cite{RegDilution}.

\begin{figure}[htbp]
            \centering
            \includegraphics[scale=0.7]{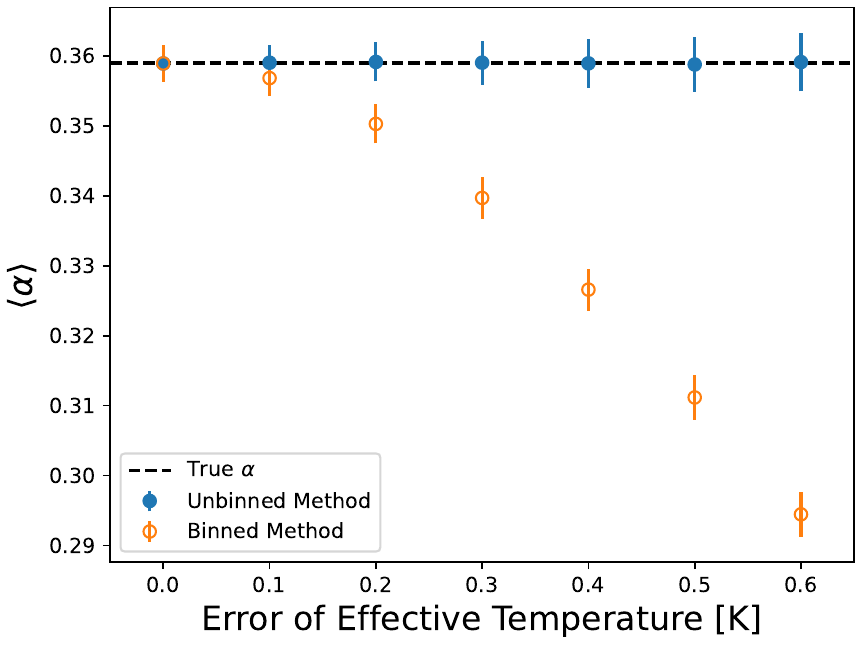}
           \caption{
           Mean correlation coefficient 
           $\left<\alpha\right>$
           as a function of  
           the effective-temperature uncertainty. Each point is obtained from a Gaussian fit to the distribution of correlation coefficients derived from 2,000 independent toyMC data sets. Results are shown for both the Unbinned Method and the Binned Method.
           }
      \label{fig:MC_result_error_linear} 
    \end{figure}

We systematically investigated the dependence of the correlation coefficient $\alpha$ on the uncertainty of  
the effective-temperature measurement $\sigma_T$. For each value of $\sigma_T$, we generated 2,000 independent toyMC data sets.  
The distribution of the resulting correlation coefficients was fit to a Gaussian distribution to extract the mean,
$\langle \alpha \rangle$,
as shown in Figure~\ref{fig:MC_result_error_linear}. Our results reveal a critical methodological divergence:
\begin{itemize}
\item{Unbinned Method:  
$\left< \alpha\right>$ 
remains stable across all $\sigma_T$ values, consistent with the theoretical prediction $\alpha_\mathrm{true}=0.359$. }
\item{Binned Method: $\left<\alpha\right>$ 
decreases  
with increasing $\sigma_T$, exhibiting a stronger negative bias in 
$\left<\alpha\right>$
as the 
effective temperature uncertainty increases.}
\end{itemize}

These results show how the Binned Method exhibits critical limitations 
in
correlation analysis involving temperature-dependent muon data. 
When the uncertainty  
in the 
independent
variable ($\Delta T_\mathrm{eff}$) is nonzero, i.e., $\sigma_T >0$, the method introduces  
nonlinear distortions and systematically underestimates the correlation coefficient. 
The conclusions remain robust under variations of the toyMC setup, including replacing the cosine modulation of the muon rate and effective temperature with alternative periodic functional forms and varying the assumed true correlation coefficient. In all tested cases, the qualitative behavior of both methods remains unchanged. 

In contrast, the Unbinned Method provides an unbiased estimate of the correlation coefficient when 
the uncertainties in the effective temperature
are properly accounted for. 
\section{Regression with effective-temperature uncertainty}\label{sec:influence}
As noted in Section~\ref{sec-UnknownTeff}, the 
uncertainty of the effective temperature is difficult to quantify.
These challenges can lead to  an incorrect estimation of $\sigma_{T}$.
To investigate the impact of such 
a misestimation on the outcome of the regression, we reused the toyMC data set with a fixed true 
uncertainty of $\sigma_T^\mathrm{true} = 0.4$~K 
and systematically assigned 
values to the $\sigma_{T_i}$
in 
Equation~\ref{eq:chisq} for each
regression 
while keeping the other simulated information unchanged. 

\begin{figure}[htbp]
            \centering
            \includegraphics[scale=0.7]{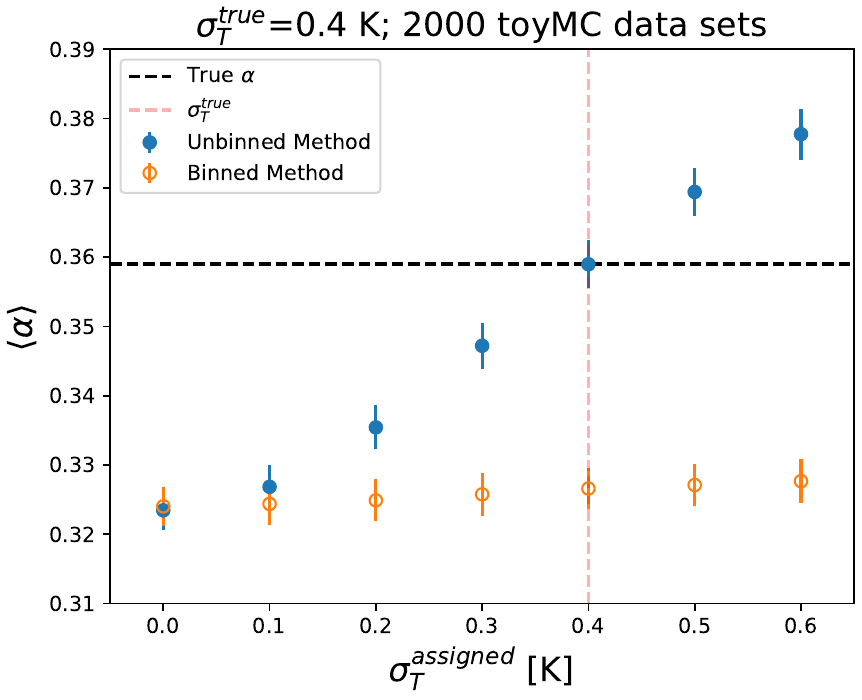}
           \caption{Correlation coefficient $\alpha$
           determined with Unbinned Method and Binned Method as a function of
           the assigned effective temperature error  
           (true uncertainty fixed at $\sigma_{T}^\mathrm{true} = 0.4$~K). 
           }
      \label{fig:MC_Fit_fixTerror} 
    \end{figure}

As shown in Figure~\ref{fig:MC_Fit_fixTerror}, the correlation coefficient $\alpha$ measured with the Unbinned Method is highly sensitive to the assigned 
effective temperature uncertainty, 
$\sigma_T^\mathrm{assigned}$.
Specifically, $\alpha$ 
increases
with $\sigma_T^\mathrm{assigned}$. When $\sigma_T^\mathrm{assigned} < \sigma_T^\mathrm{true}$, 
$\alpha$ is 
systematically smaller than $\alpha_{true}$.
If
$\sigma_T^\mathrm{assigned} > \sigma_T^\mathrm{true}$, $\alpha$ 
is larger than $\alpha_\mathrm{true}$.
On the other hand,
the Binned Method 
is insensitive
to $\sigma_T^\mathrm{assigned}$
and 
consistently underestimates the correlation coefficient $\alpha$.

This study shows that 
even the Unbinned Method  
can
produce  
a biased correlation coefficient
if
the 
uncertainty of the effective temperature 
is incorrectly estimated.

\section{Mitigation strategies for reducing bias}\label{sec:better_estimation}
 
To address the sensitivity of the Unbinned Method to mismatches between the assigned and true 
effective-temperature uncertainties identified in Section~\ref{sec:influence}, we developed a practical bias-mitigation procedure based on temporal aggregation and stability testing.

\begin{figure}[ht]
        \centering
        \includegraphics[scale=0.7]{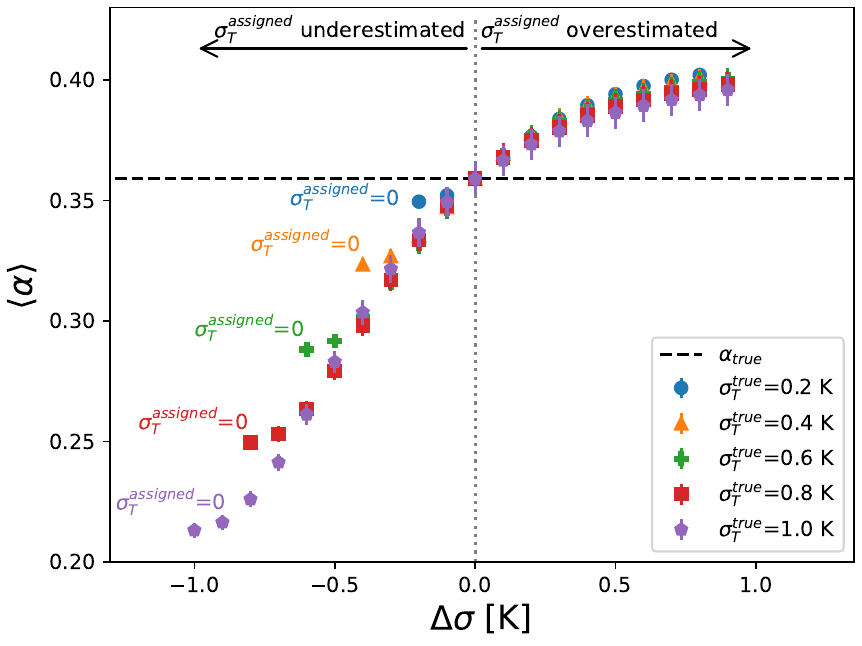}
        \caption{Correlation coefficient obtained with the Unbinned Method using toyMC samples with different true effective-temperature uncertainty $\sigma^{\mathrm{true}}_T$ as a function of the mismatch between the assigned and the true effective temperature error. Each data point is derived from 2,000 independent toyMC data sets.
       }
 \label{fig:MC_result_error_linear_changeTerror_diff_2} 
\end{figure}

First, we study the bias of the correlation coefficient as a function of the 
daily effective-temperature uncertainty mismatch, defined as: 
\begin{equation}\label{eq:error of errors}
    \Delta \sigma = \sigma^\mathrm{assigned}_{T} - \sigma^\mathrm{true}_T.
\end{equation} 
Figure~\ref{fig:MC_result_error_linear_changeTerror_diff_2} shows how the correlation coefficient from the Unbinned Method 
varies with $\Delta \sigma$ across  different $\sigma^{\mathrm{true}}_T$ levels.  Notably, the magnitude of the bias in the correlation coefficient is primarily driven by $\Delta \sigma$ and remains largely independent of $\sigma^\mathrm{true}_T$. 

These findings motivate minimizing the effective temperature uncertainty mismatch $\Delta\sigma$ as a key step to reduce
bias in the resulting correlation coefficient. 
One practical way to achieve this is to 
integrate daily data  over a longer time scale
before regression, where each data point represents an average over $n$ days. 
In our toyMC setting, the uncertainty scales as $1/\sqrt{n}$ , reducing the mismatch to: 

\begin{equation}
\label{eq:reduction}
\Delta \sigma_{n} \approx\frac{\sigma^\mathrm{assigned}_T-\sigma^\mathrm{true}_T}{\sqrt{n}} = \frac{\Delta\sigma}{\sqrt{n}}.
\end{equation}

\begin{figure}[ht]
    \centering
     \includegraphics[scale=0.7]{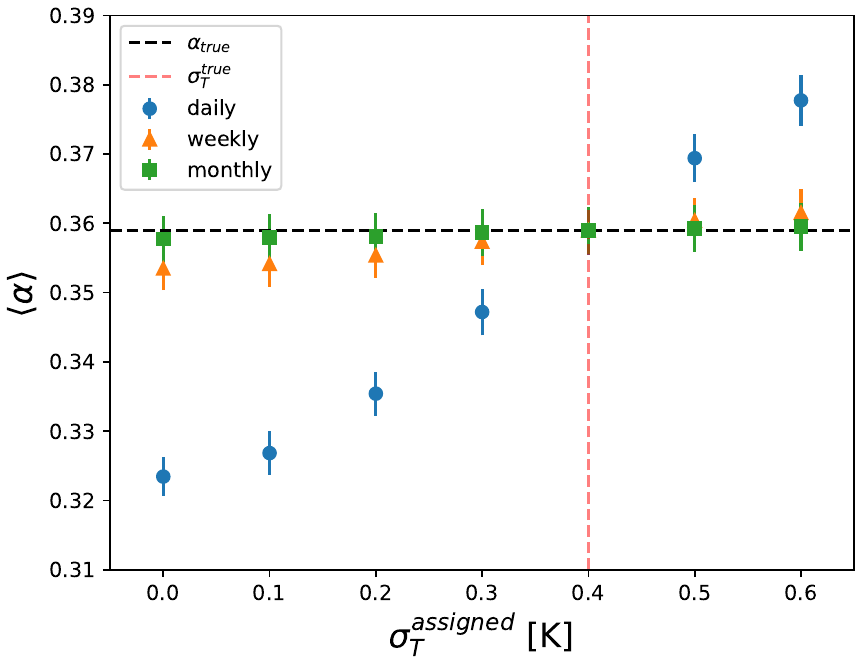}
    \caption{Correlation coefficient as a function of 
    the daily assigned effective-temperature
    uncertainty  
    for daily, weekly and monthly data sets  
    with true daily 
    effective-temperature uncertainty fixed at  $\sigma_{T}^\mathrm{true}=0.4$~K.
    Each data point is from 2,000 independent toyMC data sets.
    }
    \label{fig:toyMC_result_daily_weekly_monthly_2}
\end{figure}

To validate this approach,
we aggregated the toyMC data used in Section~\ref{sec:influence} into weekly ($n =7$) and monthly ($n =30$) intervals and performed the 
analysis again. The results, shown in Figure~\ref{fig:toyMC_result_daily_weekly_monthly_2}, exhibit a significant reduction in bias for the aggregated datasets, as expected. Even when the  
assigned effective-temperature 
uncertainty
deviates substantially from the true value, the 
best-fit
correlation coefficient approaches the true value when longer aggregation intervals are used.

\begin{figure}[!ht]
    \centering
    \includegraphics[scale=0.7]{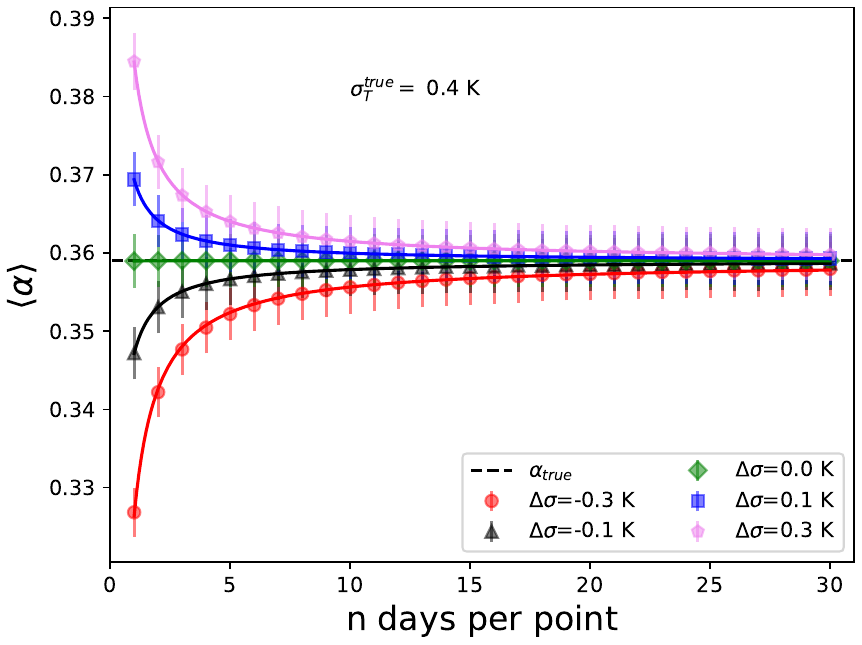}
    \caption{Correlation coefficient as a function of the number of 
    combined  
    days
    for five different daily effective-temperature uncertainty mismatches($\Delta \sigma$). The true
    uncertainty is fixed at  $\sigma^\mathrm{true}_T = 0.4$~K. 
     Each data point is derived from 2,000 independent toyMC data sets. 
    }
    \label{fig:toyMC_result_evolution}
\end{figure}

We further studied the evolution of the   
correlation coefficient as a function of the number of 
integration days  
for different 
values 
of $\Delta\sigma$.
In the context of studies of seasonal modulation of the muon rate, merging too many days  reduces
the number of data points available for regression and washes out  seasonal modulation effects. We therefore restrict our analysis to bins of  at most 30 days.
The results are shown in Figure~\ref{fig:toyMC_result_evolution}. 
We observe a systematic reduction of the bias in the correlation coefficient as the number of 
integration
days increases, regardless of the 
size of $\Delta\sigma$. 
For our toyMC setup,
the dependence is empirically well described by:
\begin{equation}
\alpha(n)=a+\frac{b}{n},
\end{equation}
where 
$a$ represents the asymptotic value of the correlation coefficient and 
$b$ characterizes the magnitude of the bias. This behavior indicates that  
combining
the data even over a modest number of days can significantly reduce the bias introduced by  
incorrect estimation of the uncertainty.

The evolution of the 
correlation coefficient as a function of the number of  
days can also serve 
as a consistency check for the 
assigned effective-temperature uncertainty. If the   
assigned uncertainty matches the true uncertainty, the 
correlation coefficient remains stable as the number of 
days increases. In contrast, incorrect uncertainty assignments lead to a systematic dependence of the 
correlation coefficient on the number of  
days, 
with the magnitude of the dependence increasing with the 
mismatch 
in the 
effective-temperature uncertainty.
The conclusion remains robust under variations of the toyMC setup, including changes to the functional form of the periodic modulation and to the input parameters. This stability criterion provides a practical method for evaluating the uncertainty of the 
effective-temperature in 
analyzing actual experimental data.

The above results were obtained using a simplified toyMC model in which all daily effective temperature measurements have identical uncertainties. However, in 
reality, measurement uncertainties  
can
vary from day to day. To investigate this scenario, we extended our study by generating toyMC data sets with daily-varying effective-temperature uncertainties. All simulation settings remain identical to those described in Section~\ref{sec-toyMC-setup}, except that the true 
effective-temperature uncertainty, $\sigma_{T}^\mathrm{true}$, for each day is sampled from a Gaussian distribution 
with a mean of 0.4~K and standard deviation of 0.05~K, 
$\mathcal{N} (0.4~\mathrm{K}, 
(0.05~\mathrm{K})^2)$.

\begin{figure}[!ht]
    \centering
    \includegraphics[scale=0.7]{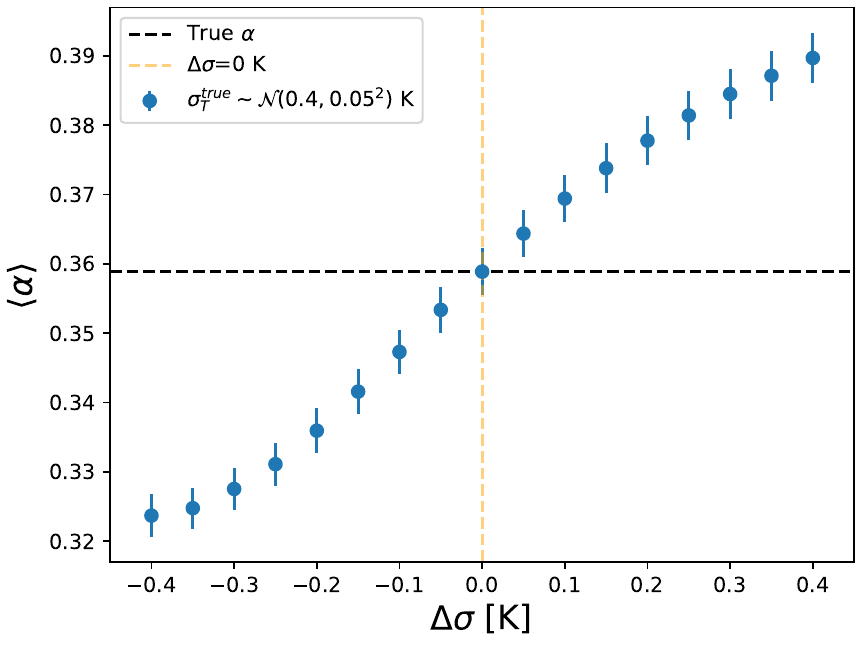}
    \caption{Correlation coefficient as a function of 
   the daily effective-temperature uncertainty mismatch ($\Delta\sigma$). 
    In these simulations,
    true daily effective-temperature
    uncertainty varies  and is draw from a Gaussian distribution. The  
    daily   assigned 
    effective-temperature
    uncertainty used in the fitting process is set to this true value plus the specific 
    mismatch 
    $\Delta \sigma$.  
     Each data point is
    derived from 2,000 independent toyMC data sets.
    }
    \label{fig:VarTerr_Dsigma}
\end{figure}

The dominant factor that controls the bias remains the 
uncertainty mismatch
$\Delta\sigma$, as illustrated in Figure~\ref{fig:VarTerr_Dsigma}. In this test,  
the assigned 
daily effective-temperature
uncertainty 
follows the true day-to-day variations 
of $\sigma_{T}^\mathrm{true}$.
When the assigned 
daily effective-temperature
uncertainties accurately match the 
corresponding
true 
daily
uncertainties, the Unbinned Method successfully recovers the true correlation coefficient.

\begin{figure}[!ht]
    \centering
    \includegraphics[scale=0.7]{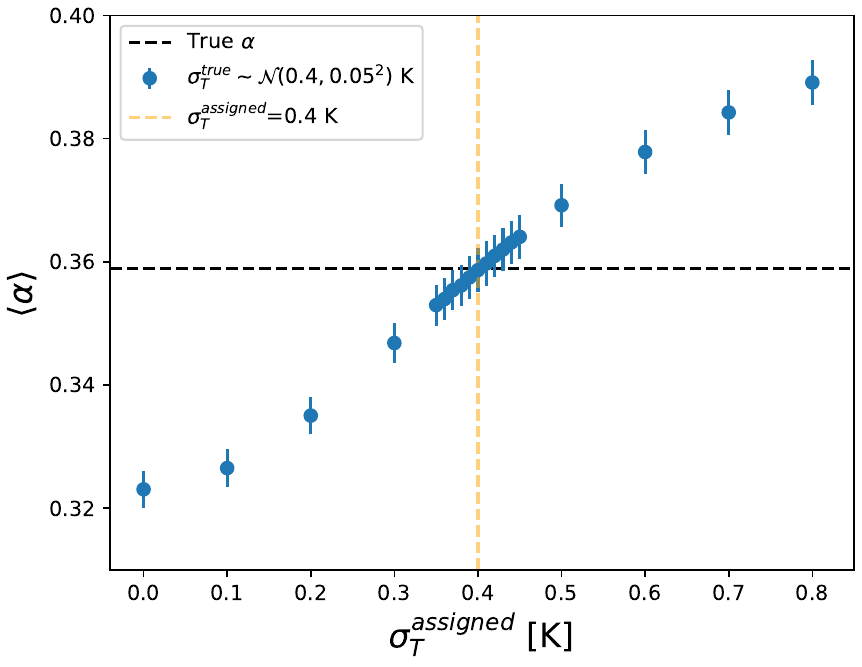}
    
    \caption{ Dependence of the  
    correlation coefficient on the 
    fixed
    assigned effective-temperature uncertainty $\sigma^\mathrm{assigned}_T$. In these simulations, the daily true 
    effective-temperature uncertainty varies and is drawn from a Gaussian distribution.
Each data point is 
derived from 2,000 independent toyMC data sets.
    }
    \label{fig:VarTerr_constantTerr_2}
\end{figure}

We also explored a second scenario in which a single constant uncertainty $\sigma_T^\mathrm{assigned}$ is used in the regression, independent of the day-to-day variations of $\sigma_T^\mathrm{true}$. The results are shown in Figure~\ref{fig:VarTerr_constantTerr_2}. In this case, the bias in the estimated correlation coefficient becomes negligible when the assigned constant uncertainty is close to the mean value of the true uncertainty distribution. We tested several distributions for the daily effective temperature uncertainty and found that the mean of the distribution is a robust choice for the constant assigned uncertainty. This suggests that in situations where detailed day-to-day uncertainty estimates are unavailable, assigning a constant effective temperature uncertainty close to the typical uncertainty scale can still provide an approximately unbiased estimate of the correlation coefficient.

\begin{figure}[!ht]
    \centering
    \includegraphics[scale=0.7]{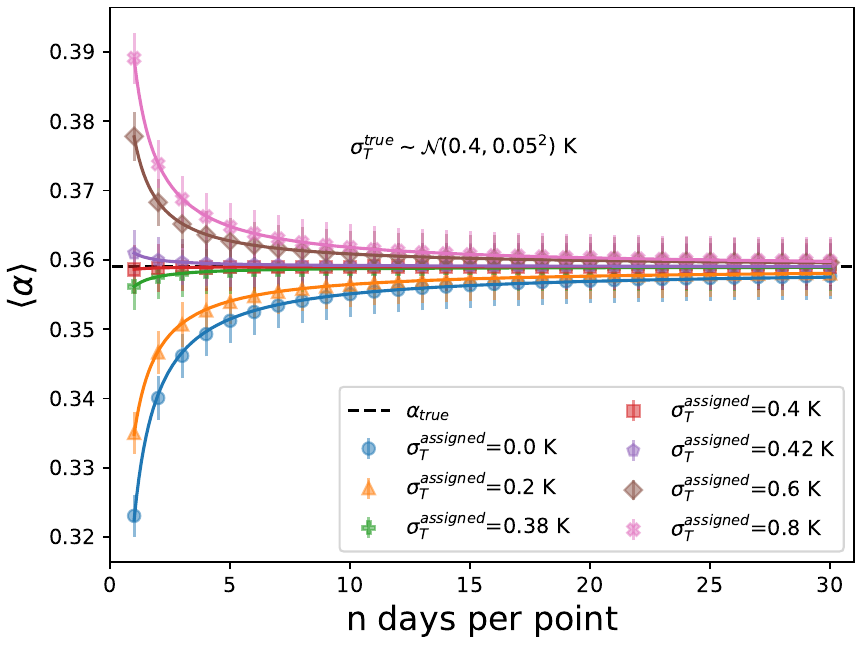}
    \caption{Correlation coefficient as a function of the number of 
    sampling
    days
    for seven fixed assigned effective-temperature uncertainties.  
    The daily true 
    effective-
    temperature uncertainty varies, drawn from a Gaussian distribution. 
    Each data point is
    derived from 2,000 independent toyMC data sets. 
    }
    \label{fig:VarTerr_days}
\end{figure}

This  result has  direct practical value. 
In the studies of seasonal modulation of the muon rate,
one can assign a constant error and test whether the inferred correlation coefficient is stable across time aggregations. It also serves as a robust alternative when the true daily uncertainty is unknown or cannot be reliably estimated. A demonstration of this 
technique
is shown in Figure~\ref{fig:VarTerr_days}, where a constant 
$\sigma^\mathrm{assigned}_T$
was used to 
determine the
correlation coefficients from the data with daily-varying 
effective-temperature uncertainty. The mean of the 
distribution of the daily effective-temperature uncertainties 
provides a good initial estimate of $\sigma_T^\mathrm{assigned}=0.4$~K, resulting in  essentially no change in the 
correlation
coefficient 
regardless of  
the number of days combined.

\section{Discussion and conclusion}\label{sec:conclusion}

    This study used a toyMC simulation as a controlled environment in which to assess the performance of the Unbinned and Binned Methods commonly used to derive the correlation between the underground muon rate and the effective atmospheric temperature. The simulation assumed a perfect linear relationship between the muon rate and the effective temperature and explored different values of the temperature uncertainties. 
    The toyMC parameters were chosen to reproduce experimentally realistic amplitudes and uncertainty scales observed in underground muon measurements. We find that, while both methods yield correct results in the absence of temperature uncertainties, the Binned Method becomes biased when temperature uncertainties are non-zero. Moreover, the Unbinned Method is highly sensitive to the magnitude of the assumed temperature uncertainties, returning the correct value of the correlation coefficient only when they are properly estimated.
    
    In light of these results, we propose that muon modulation analyses adopt the \textbf{Unbinned Method} as the standard approach for correlation studies involving effective temperature, provided that realistic uncertainties in $T_\mathrm{eff}$ are used. When such uncertainties are poorly constrained, the bias can be mitigated by calibrating the error model via the stability of the inferred correlation coefficient $\alpha$ with respect to temporal binning, namely, by identifying the error assignment that renders the correlation coefficient insensitive to the number of days. In other words,  if $\alpha$ remains unchanged as the data is aggregated over increasingly longer time intervals, then 
    the assigned uncertainty of $T_\mathrm{eff}$ represents the correct uncertainty
    estimate and subsequently yields an unbiased value of $\alpha$. 
    This procedure remains robust even in the presence of temporal variations in the true effective temperature uncertainty, provided that the single constant uncertainty used in the analysis is close to the mean of the true distribution.

    We emphasize that our analysis accounts only for uncorrelated random errors in both the muon rate and effective temperature measurements. Correlated systematic uncertainties 
    affecting the muon rate, the effective temperature, or both, are beyond the scope of this study.

\section*{Acknowledgements}
The authors thank the Daya Bay Collaborators for their support and valuable discussions. We are especially grateful to Maxim  Gonchar and David Jaffe for encouraging this study. The work of Bangzheng Ma and Qun Wu was supported by the National Natural Science Foundation of China under Grant No. 11875180. The work of Bed\v{r}ich Roskovec was supported by the Charles University Center of Excellence UNCE/24/SCI/016. Kam-Biu Luk was partially supported by the U.S. Department of Energy under contract number OHEP DE-AC02-05CH11231, the Paul and May Chu Foundation, and the HK JEBN Limited. The work of Juan Pedro Ochoa-Ricoux and Katherine Dugas was supported by the U.S. Department of Energy, Office of Science, Office of High Energy Physics, under Award Number DE-SC0009920.




\bibliographystyle{JHEP}
\bibliography{ref}

\end{document}